\documentclass[12pt,preprint]{aastex}

\slugcomment{ApJ in press}

\shorttitle{XMM-Newton Observations of A133}
\shortauthors{Fujita et al.}

\begin{document}

\title{{\it XMM-Newton} Observations of A133: A Weak Shock Passing
through the Cool Core}

\author{Yutaka Fujita\altaffilmark{1,2},
Craig L. Sarazin\altaffilmark{3},
Thomas H. Reiprich\altaffilmark{3},
H. Andernach\altaffilmark{4},
M. Ehle\altaffilmark{5},
M. Murgia\altaffilmark{6,7},
L. Rudnick\altaffilmark{8},
and
O. B. Slee\altaffilmark{9}
}

\altaffiltext{1}{National Astronomical Observatory, Osawa 2-21-1,
Mitaka, Tokyo 181-8588, Japan; yfujita@th.nao.ac.jp}
\email{yfujita@th.nao.ac.jp}

\altaffiltext{2}{Department
of Astronomical Science, The Graduate University for Advanced Studies,
Osawa 2-21-1, Mitaka, Tokyo 181-8588, Japan}

\altaffiltext{3}{Department of Astronomy, University of
Virginia, P. O. Box 3818, Charlottesville, VA 22903-0818, USA;
sarazin@virginia.edu, thomas@reiprich.net}

\altaffiltext{4}{Depto.\ de Astronom\'\i a, Univ.\ Guanajuato, Apdo.\
Postal 144, Guanajuato, C. P.\ 36000, GTO, Mexico; heinz@astro.ugto.mx}

\altaffiltext{5}{XMM-Newton Science Operations Centre, European Space
   Agency, Villafranca del Castillo, P.O.Box 50727, 28080 Madrid, Spain;
   mehle@xmm.vilspa.esa.es}

\altaffiltext{6}{Istituto di Radioastronomia (CNR), Via Gobetti 101,
I-40129 Bologna, Italy; murgia@ira.bo.cnr.it}

\altaffiltext{7}{Osservatorio Astronomico di Cagliari, Loc. Poggio dei
Pini, Strada 54, I-09012 Capoterra (CA), Italy}

\altaffiltext{8}{Department of Astronomy, University of Minnesota, 116
Church Street SE, Minneapolis, MN 55455; larry@astro.umn.edu}

\altaffiltext{9}{Australia Telescope National Facility, CSIRO, PO Box 76,
Epping, NSW 1710, Australia; Bruce.Slee@atnf.csiro.au}

\begin{abstract}
We use {\it XMM-Newton} observations of the cluster of galaxies A133 to
study the X-ray spectrum of the intracluster medium (ICM). We find a
cold front to the southeast of the cluster core. From the pressure
profile near the cold front, we derive an upper limit to the velocity of
the core relative to the rest of the cluster of $< 230$ km s$^{-1}$.
Our previous {\it Chandra} image of A133 showed a complex, bird-like
morphology in the cluster core.  Based on the {\it XMM-Newton} spectra
and hardness ratio maps, we argue that the wings of this structure are a
weak shock front. The shock was probably formed outside the core of the
cluster, and may be heating the cluster core.  Our {\it Chandra} image
also showed a ``tongue'' of relatively cool gas extending from the
center of the cD to the center of the radio relic. The {\it XMM-Newton}
results are consistent with the idea that the tongue is the gas which
has been uplifted by a buoyant radio bubble including the radio relic to
the northwest of the core.  Alternatively, the tongue might result from
a cluster merger. The small velocity of the core suggests that the
bubble including the relic has moved by buoyancy, rather than by motions
of the core or the ICM. We do not find clear evidence for nonthermal
X-ray emission from the radio relic.  Based on the upper limit on the
inverse Compton emission, we derive a lower limit on the magnetic field
in the relic of $B\geq 1.5$~$\mu$G.
\end{abstract}


\section{Introduction}

A133 is an X-ray luminous cluster at $z=0.0562$ \citep*{way97}. The
cluster shows substructure which may indicate that it is undergoing a
merger \citep*{kry99}. However, it was said that there is a moderate
cooling flow at the center of the cluster \citep*[$\dot{M} = 110 \,
M_\sun\: \rm yr^{-1}$ for $H_0=50\:\rm km\: s^{-1}\:
Mpc^{-1}$;][]{whi97}.  The central cD galaxy is a radio source.  About
30$''$ north of the cD galaxy, there is a diffuse radio relic source,
which has a ``spidery'' filamentary structure and an extremely steep
spectral index \citep{sle01}. {\it ROSAT} observations suggested that
there might be excess X-ray emission associated with the radio relic
\citep*{sle01,riz00}.

\citet{fuj02} observed A133 with {\it Chandra} and showed that the
central cool core is irregular; the dominant feature is a ``tongue'' of
cool X-ray gas which extends to the northwest of the cD galaxy. The
tongue ends on and overlaps the radio relic, at least in
projection. Excluding the tongue, the X-ray surface brightness at the
relic is about 30\% fainter than the surrounding region.  There is no
clear evidence for a merger shock near the radio relic, at least in the
{\it Chandra} image. On the other hand, the {\it Chandra} spectrum gave
a (at best) marginal detection of inverse Compton emission. The position
of the cD galaxy coincides with the brightest X-ray emission, but is
shifted to the southeast of the X-ray centroid determined on larger
scales in the cluster.

\citet{fuj02} discussed several possible origins for the tongue; two of
these seem reasonably consistent with the X-ray and radio data. First,
the tongue may be the result of a Kelvin-Helmholtz instability acting on
the cool core surrounding the cD galaxy as the cD galaxy moves through
surrounding, lower density gas. The second possibility is that the radio
relic is a buoyant radio bubble and that the tongue was uplifted by the
motion of this bubble, as appears to have occurred in M87/Virgo.  In
order to constrain these models further, detailed spectral information
is needed.

In this paper, we report the {\it XMM-Newton} observations of A133.
{\it XMM-Newton} has a larger 
field of view and much more collecting area at high energies than {\it
Chandra}.  Combining these rich spectral data with the existing {\it
Chandra} images, we discuss the origin of the complicated central
structures of A133.  We assume $H_0=70\:\rm km\: s^{-1}\: Mpc^{-1}$,
$\Omega_0=0.3$, and $\Lambda=0.7$ unless otherwise noted.  At a redshift
of 0.0562, $1\arcsec$ corresponds to 1.07~kpc.

\section{{\it XMM-Newton} Observation and Data Analysis}

A133 was observed with {\it XMM-Newton} \citep{jan01} on 2002 December
22--23 for a total exposure time of 33.7~ks.  The EPIC-MOS cameras were
operated in Full Frame mode and the EPIC-PN camera in Extended Full
Frame mode \citep{tur01,str01}. The Thin filter was used in all EPIC
observations. We created calibrated event files using SAS Version 5.4.1,
and all further reduction and analysis used this version of SAS.

Data obtained with {\it XMM-Newton} are often affected by periods of
high background flares, which need to be removed. We adopted the
cleaning method described in detail in \citet{rei04}. In order to
determine lightcurves free from the contamination from astrophysical
sources, we chose the energy band of $10$--$12$~keV and $12$--$14$~keV
for the MOS and PN cameras, respectively.  The exposure time was binned
in 100~s intervals and only events with pattern $\le 12$ and
$\sharp$XMMEA\_EM (MOS) or flag $=0$ (PN) were used. We applied
conservative `generic' cuts to the count rates $C_X$ before we estimated
the mean count rates to find background flares. For MOS (PN), only times
with $0.01\:(0.02)\le C_X \le 0.25\:(0.45)$ counts per second have been
used. Finally, we obtained the good time intervals (GTIs) by including
only times where the count rates are within $\pm 2\sigma$ of the mean
count rates. The final count-rate ranges are $0.03$--$0.17$ (MOS1),
$0.02$--$0.18$ (MOS2), and $0.09$--$0.30$ (PN) counts per second. The
useful exposure times after these procedures are 20.9~ks (MOS1 and MOS2)
and 15.4~ks (PN).

After the flaring background is removed, we further need to remove the
quiet background. Since the emission from the cluster covers most of the
fields of view of the detectors, we used the quiet background created
from blank sky observations\footnote{The files are available at
http://xmm.vilspa.esa.es/external/xmm\_sw\_cal/calib/epic\_files.shtml}.
These background files have been screened for high flaring background
periods and background spectra have been extracted with the same
selection criteria as the source spectra.  We found source-to-background
count-rate ratios of 1.10 for MOS1, 1.09 for MOS2, and 1.17 for the PN
detector and rescaled the background files by rewriting the BACKSCAL
header keyword according to these values. We note that the results are
not much affected by quiet background, because we mostly focus on the
bright central region of the cluster.

\begin{figure}\epsscale{0.9}
\plottwo{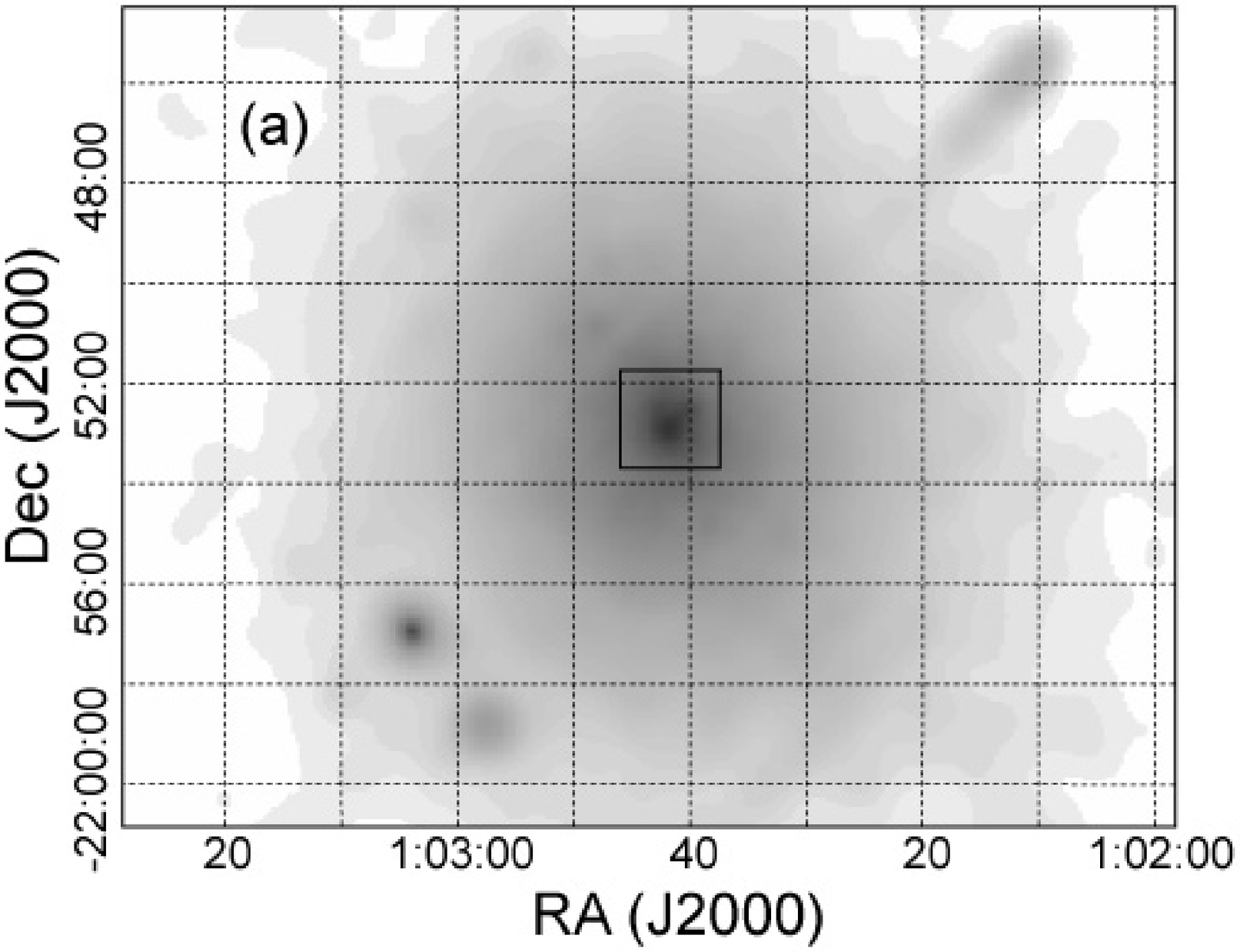}{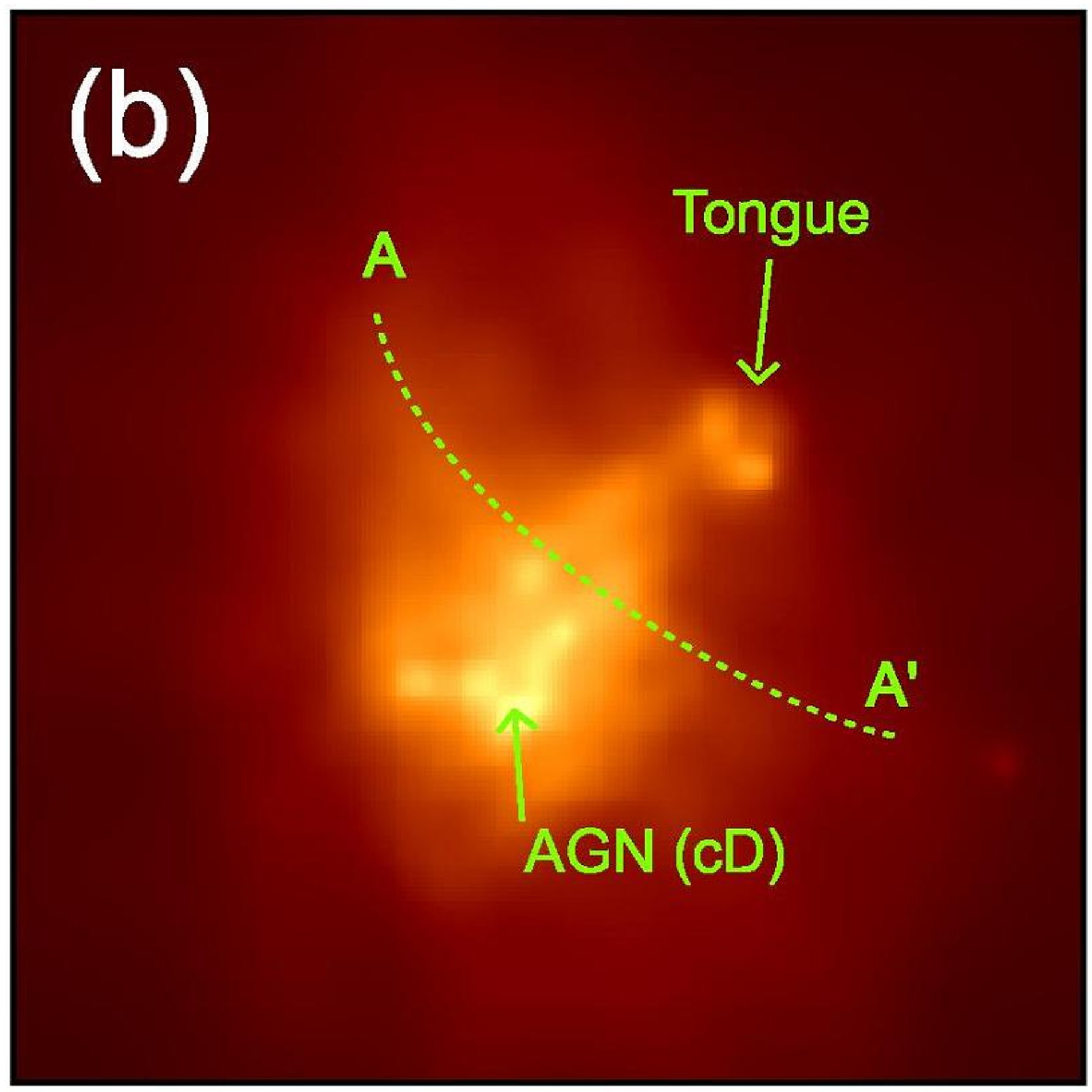}
\caption{(a) Adaptively smoothed, combined
MOS1-MOS2-PN image of A133 in the energy range of $0.3-2$~keV, corrected
for background, exposure, and vignetting.  North is up and east is
left. The linear artifacts are due to an imperfect exposure correction
near the chip boundaries. (b) Adaptively smoothed {\it Chandra} image of
the central $2'\times 2'$ region of A133, corrected for background,
exposure, and vignetting.  The region shown in Figure~\ref{fig:image}b
is indicated as a solid-line square in Figure~\ref{fig:image}a.  The
curve AA$'$ is a possible weak shock (see
\S\ref{sec:spectra_shock}). The tongue and AGN are indicated by arrows.
\label{fig:image}}
\end{figure}

Figure~\ref{fig:image}a shows the combined MOS1-MOS2-PN image of A133 in
the $0.3$--$2$~keV energy band; this band was used to emphasize the
complicated central structure of the cluster, which is mainly due to gas
being cooler and thus more prominent at low energies. The image is
corrected for background, exposure, and vignetting, and was adaptively
smoothed to a minimum signal-to-noise ratio of 15 per smoothing beam.
For comparison, in Figure~\ref{fig:image}b we also present the central
$2\arcmin\times 2\arcmin$ image obtained with {\it Chandra}
\citep[Fig.~1b in ][]{fuj02}. Figure~\ref{fig:hardness} shows the
hardness ratio map of A133.  We define the hardness ratio as the
combined MOS1-MOS2-PN count rate in the $2$--$10$~keV band divided by
the count rate in the $0.3$--$2$~keV band. The central region and the
tongue are cold. There is a temperature jump at the southeast edge of
the cool core, which will be discussed in \S~\ref{sec:spectra_projected}
and \ref{sec:coldfront}.

\section{Spectral Analysis}
\label{sec:spectra}

We extracted spectra in the $0.5$--$10$~keV (MOS) and $0.3$--$12$~keV
(PN) bands from selected regions of the {\it XMM-Newton} image using the
SAS software package. Only single and double events were used to extract
PN spectra. The response matrices we used were
m11\_r7\_im\_all\_2002-11-07.rmf for MOS1,
m21\_r7\_im\_all\_2002-11-07.rmf for MOS2, and epn\_ef20\_sdY9.rmf for
PN.  The ancillary response files were calculated using {\tt arfgen}
(Version 1.48.8). The spectra were grouped to have a minimum of 50
counts per bin and fitted with spectral models using
XSPEC\footnote{http://heasarc.gsfc.nasa.gov/docs/xanadu/xspec/}, Version
11.2. All errors are quoted at the 90\% confidence level unless
otherwise mentioned.  Spectra obtained with MOS1 and MOS2 were fitted
simultaneously; fitting MOS1 and MOS2 separately gives consistent
results.

\subsection{Projected Profiles}
\label{sec:spectra_projected}

Figures~\ref{fig:temp_proj} and~\ref{fig:abun_proj} show the radial
profiles of projected temperature and abundance for four sectors,
respectively. We define the projected temperature (projected abundance)
to be the X-ray emission-weighted temperature (abundance) along the line
of sight. Spectra obtained with MOS1 and MOS2 were simultaneously
fitted. The position angles (north through east) are
$0^{\circ}$--$90^{\circ}$ (north east; NE), $90^{\circ}$--$180^{\circ}$
(south east; SE), $180^{\circ}$--$270^{\circ}$ (south west; SW), and
$270^{\circ}$--$360^{\circ}$ (north west; NW).  The complexity of the
central structure of the cluster makes it difficult to select a center
for the sectors.  The center that we selected ($01^{\rm h}02^{\rm
m}41\fs6$; $-21\degr52\arcmin49\arcsec$, J2000) is the centroid of the
bright core of the X-ray emission, which we took to be outlined by the
X-ray brightness contour having a radius of $\sim 20\arcsec$ in the
central region of the cluster.  The spectra were fitted by a single
thermal model (MEKAL) with variable absorption (WABS).  For the NW
sector, the spectra in Figures~\ref{fig:temp_proj}
and~\ref{fig:abun_proj} do not include photons from the X-ray tongue; we
excluded the emission of the $30^\circ$ sector centered on the tongue
for $24\arcsec <r< 49 \arcsec$.  The temperature and abundance of the
tongue region are shown in Figure~\ref{fig:temp_ton_proj}. We selected
the spatial bin sizes so that errors of temperatures are $\sim 15$\%,
which corresponds to photon counts of $\sim 2000$ for each MOS and $\sim
4000$ for PN. Bin sizes can be regarded as a 'spatial resolution' for
the spectral analysis.

\begin{figure}\epsscale{0.80}
\plotone{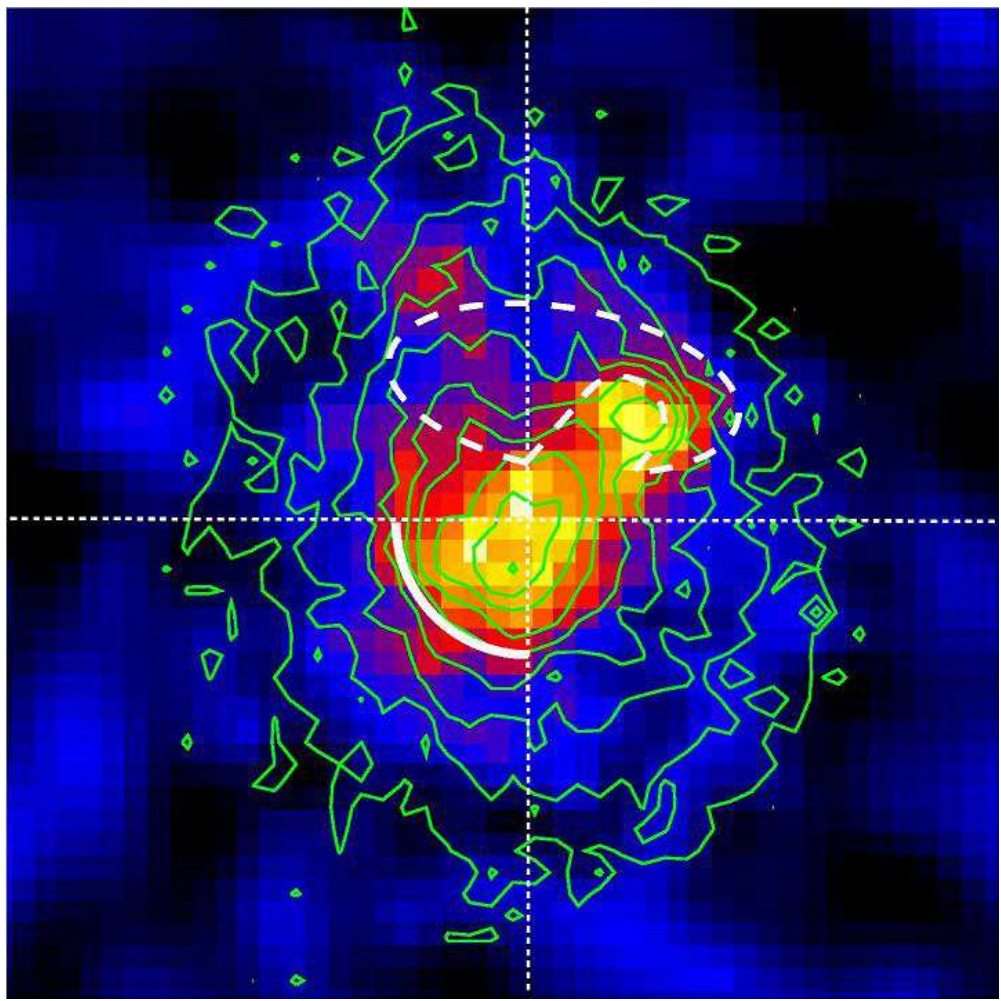} \caption{Color map of hardness ratio of the central
 $3'\times 3'$ region of A133, based on the ratio of the combined
 MOS1-MOS2-PN count rates in the energy bands $0.3$--$2.0$ keV~and
 $2.0$--$10$~keV. The map has been corrected for background, exposure,
 and vignetting, and is adaptively smoothed. Soft emission appears
 yellow and hard emission black. Also shown are surface brightness
 contours from {\it Chandra} (see Fig.~\ref{fig:image}b). The thin white
 dotted lines show the 4 quadrants used for the analysis of the profiles
 of temperature and abundance (\S\ref{sec:spectra_projected} \&
 \ref{sec:spectra_deprojected}). The dashed line shows the radio relic
 region (see \S\ref{sec:spectra_relic}). The thick white solid line in
 the SE sector is a cold front (see \S\ref{sec:coldfront}).
 \label{fig:hardness}}
\end{figure}

\begin{figure}\epsscale{1.0}
\plotone{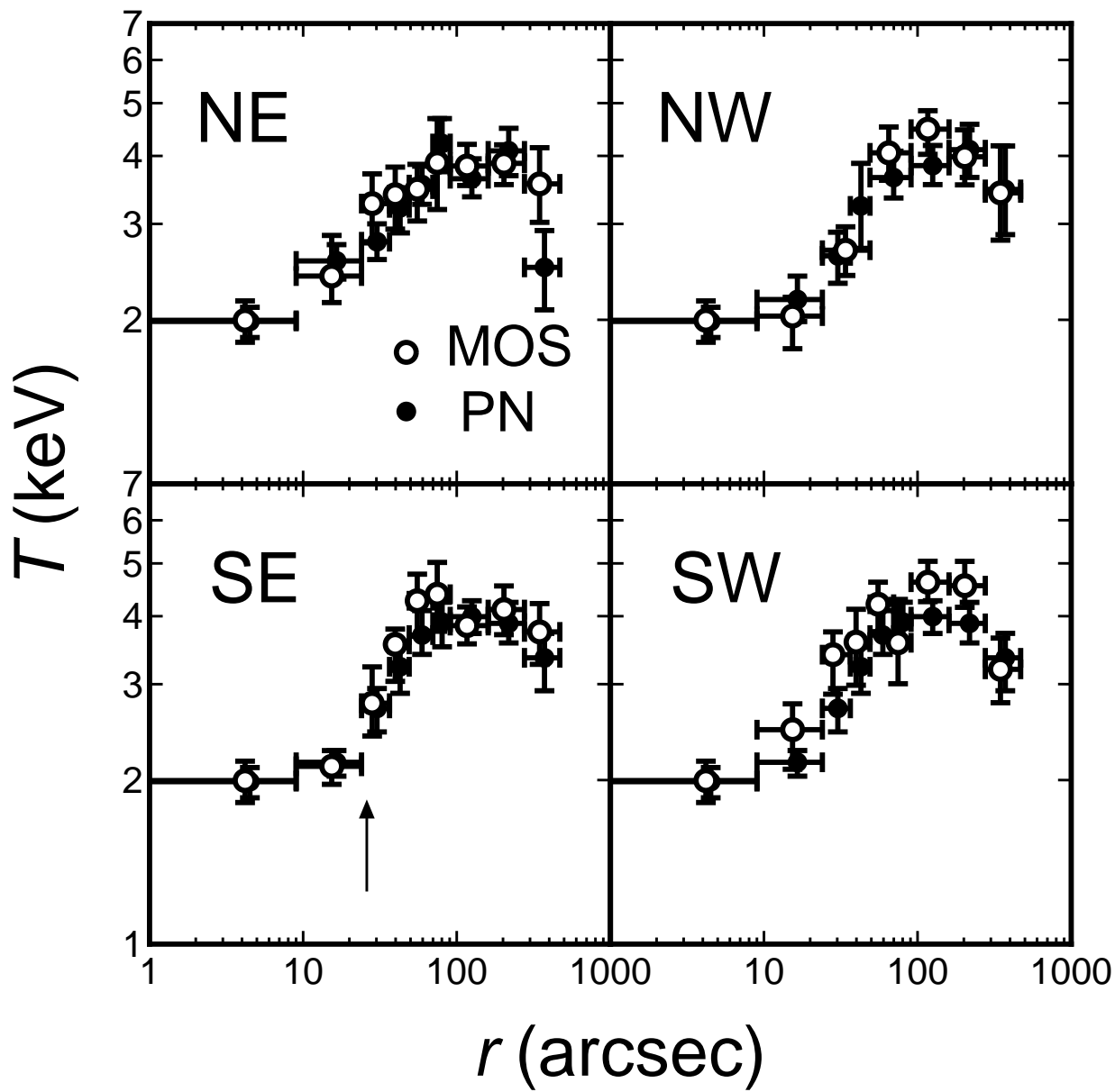} \caption{Projected temperature as a function of radius
 from the {\it XMM-Newton} observation.  Temperatures obtained from the
 two MOS detectors are shown as open circles, while temperatures
 obtained from the PN are shown as solid circles.  A temperature jump
 corresponding to a cold front is shown by an arrow in the SE
 panel. \label{fig:temp_proj}}
\end{figure}

\begin{figure}\epsscale{1.0}
\plotone{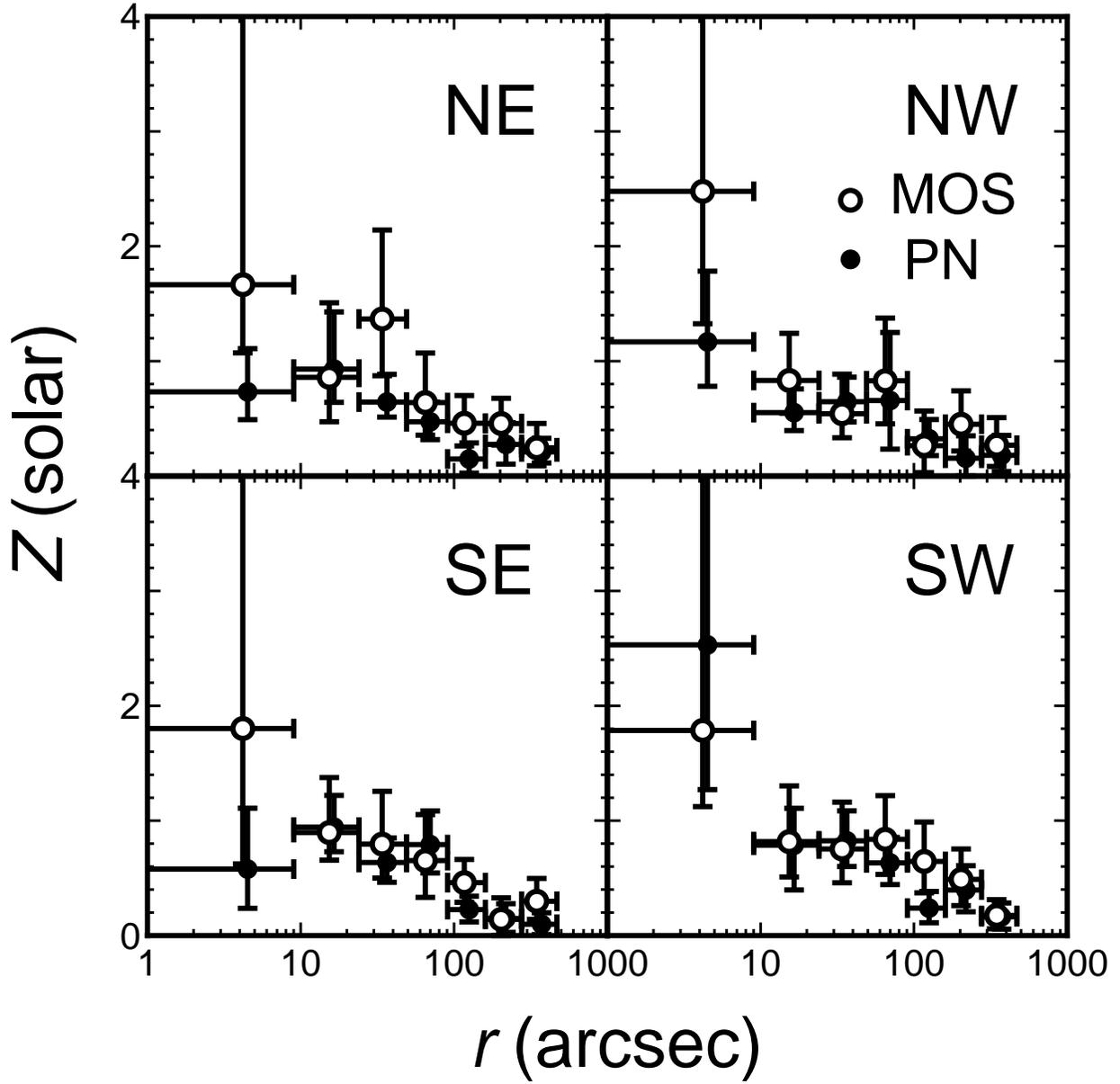} \caption{Projected metal abundance as a function of
 radius in the {\it XMM-Newton} observation. Abundances obtained from
 MOS are shown as open circles, and abundances obtained from PN are
 shown as solid circles. \label{fig:abun_proj}}
\end{figure}

\begin{figure}\epsscale{1.0}
\plottwo{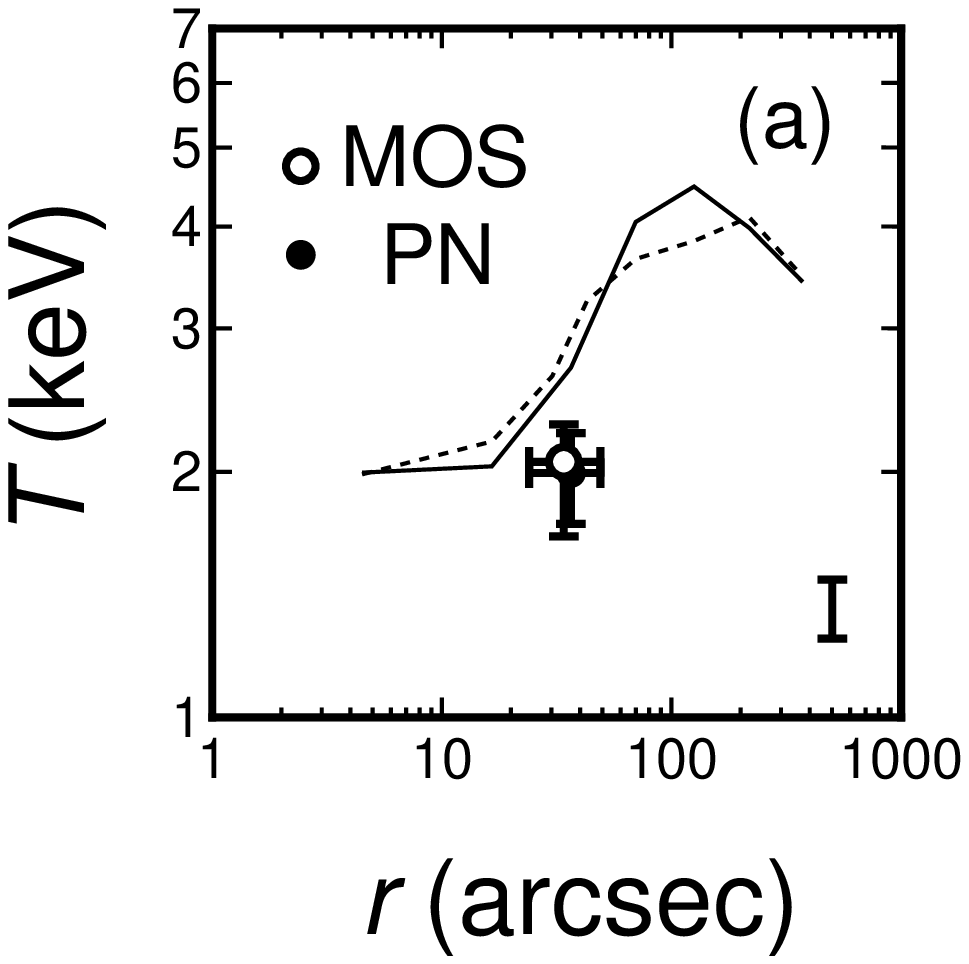}{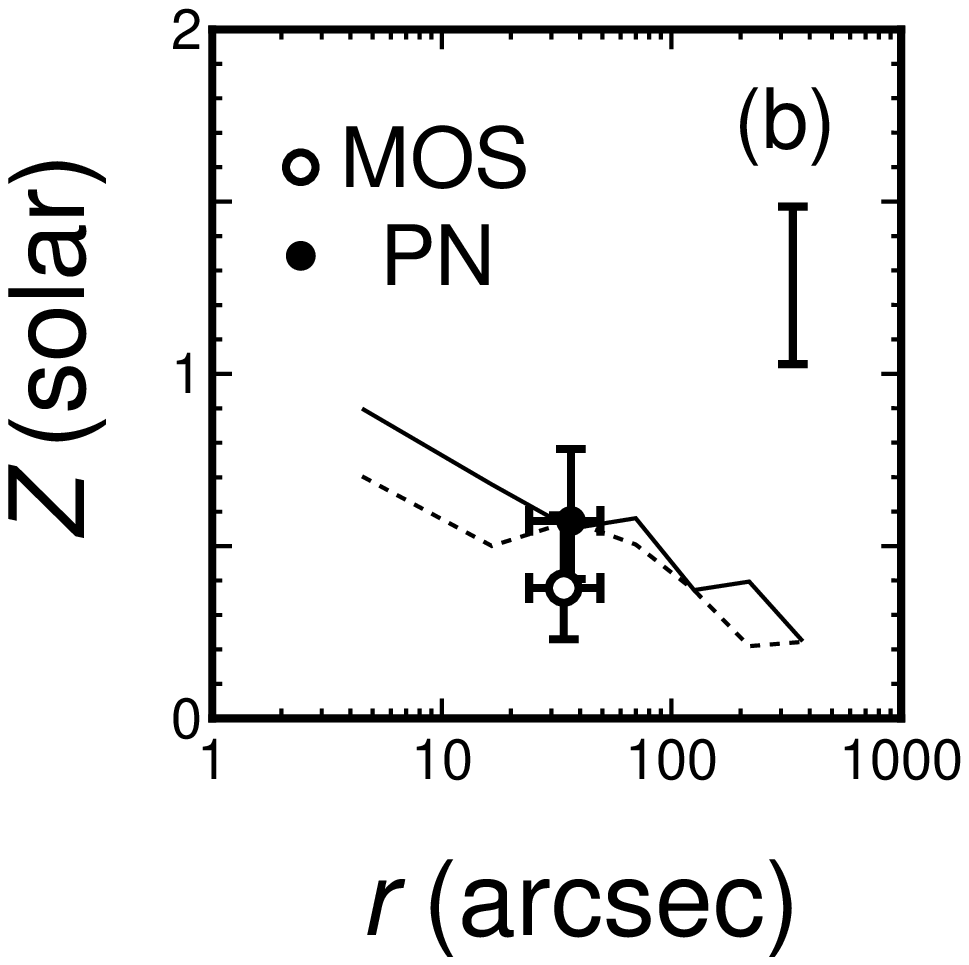} \caption{(a) Projected temperature of the
 tongue. A temperature obtained from MOS is shown as an open circle; the
 temperature from PN is almost exactly the same, and this point lies
 beneath the MOS point. The solid line shows the best fit temperatures
 obtained from MOS in the NW sector shown in Figure~\ref{fig:temp_proj},
 excluding the emission from the tongue. The dotted line is that
 obtained from PN. The typical error for the lines are shown by the bar
 at the lower right corner. (b) The same as Figure~(a) but for metal
 abundance.  \label{fig:temp_ton_proj}}
\end{figure}

In general, temperature increases and abundance decreases outward
(Figs.~\ref{fig:temp_proj} and~\ref{fig:abun_proj}). We found a
temperature jump at $r\approx 24\arcsec$ in the SE sector
(Fig.~\ref{fig:temp_proj}). Here, we define a jump as a temperature
gradient between two adjacent points where total error bars, which are
obtained by simply joining the MOS and PN error bars at each point, do
not overlap. Although we call this temperature gap a jump, the upper
limit of the width is $\sim 25$~kpc
(Fig.~\ref{fig:temp_proj}). Moreover, we do not deny the existence of
smaller jumps in other regions that cannot be resolved with the current
spatial resolution for the spectral analysis. However, the jump we found
in the SE direction appears to be significant, because it is also seen
in the hardness ratio map (Fig.~\ref{fig:hardness}) and it also
corresponds to the jump in the surface brightness profile found in the
{\it Chandra} image, which is indicated as `jump' in Figure~6 in
\citet{fuj02}. Since the width of the jump in the hardness ratio map and
in the surface brightness profile is much smaller than 25~kpc, it is
likely that the width of the temperature jump is also smaller than
25~kpc. Figure~\ref{fig:temp_proj} also shows that there are relatively
steep temperature gradients at $r\approx 24\arcsec$ in the NW and SW
sectors. Since the surface brightness profiles are also relatively steep
there (Fig.~\ref{fig:hardness}), the jump in the SE sector may continue
on into the SW and NW sectors. The temperature of the tongue is
significantly smaller than that of the surrounding ICM
(Fig.~\ref{fig:temp_ton_proj}a), a result already found from {\it
Chandra} data \citep{fuj02}. The abundance of the tongue is almost the
same as that of the surrounding ICM (Fig.~\ref{fig:temp_ton_proj}b).

\begin{figure}\epsscale{0.45}
\plotone{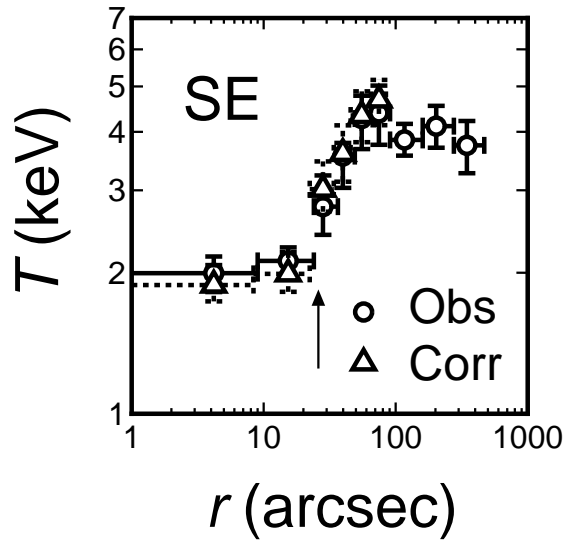} \caption{Projected temperature as a function of radius
 for the SE sector for MOS. Open circles are the same as those in
 Figure~\ref{fig:temp_proj}. Open triangles are the temperature profile
 corrected for the PSF. \label{fig:temp_proj_psf}}
\end{figure}

Note that in Figures~\ref{fig:temp_proj} and~\ref{fig:abun_proj}, the
widths of the radial bins are comparable to the point spread function
(PSF) of {\it XMM-Newton} ($\sim 6''$ for the full width half maximum,
FWHM, and $\sim 15''$ for the half energy width, HEW) in the central
region. Therefore, we investigated the influence of the PSF on the
derived temperature profiles. We choose the SE sector, because the
temperature gradient is the steepest among the four sectors and it
should be affected by the PSF most significantly. We assumed that the
PSF is constant across the extent of the cluster and that its shape is
energy independent following \citet*{maj02}. We convolved the surface
brightness profile obtained by {\it Chandra}, which has a much better
angular resolution ($0.5''$) than {\it XMM-Newton}, and the PSF of {\it
XMM-Newton} \citep{ghi01}. We then estimated the contribution of flux to
a given radial bin from other bins. Weighting with the fluxes, we
calculated the corrected temperature profile that can reproduce the
observed temperature profile. The result is shown in
Figure~\ref{fig:temp_proj_psf} for MOS. The differences between the
corrected and observed profiles are very small; error bars are well
overlapped. This is because A133 does not have a strong X-ray peak at
the center and the angular size of the cool core is larger than the PSF
of {\it XMM-Newton}. In addition, the spectrum of the tongue emission is
not affected by the central emission because the tongue is bright
enough. For the azimuthal direction, the gradient of the surface
brightness is smaller than that in the radial direction for the SE
sector except for the periphery of the tongue. Therefore, we ignore the
influence of the PSF from now on except for the periphery of the tongue.

\subsection{Deprojection Analysis}
\label{sec:spectra_deprojected}

The above radial profiles of temperature and abundance are affected by
projection effects. This problem can be solved by deprojection
analysis. We used the same deprojection procedure as that of
\citet*{bla03}.  The spectrum from the outermost annulus was fitted with
a single-temperature MEKAL model with the absorption fixed to the
Galactic value \citep[$1.58\times 10^{20}\rm\:
cm^{-2}$;][]{sta92}. Then, the next annulus inside was fitted. The model
used for this annulus was a combination of the best-fitting model of the
exterior annulus with the normalization scaled to account for the
spherical projection of the exterior shell onto the inner one, along
with another MEKAL component added to account for the emission at the
radius of interest. This process was continued inward, and we fitted
one spectrum at a time. As is the case of projected profiles, we
exclude the emission from the tongue for the NW sector.

\begin{figure}\epsscale{1.0}
\plotone{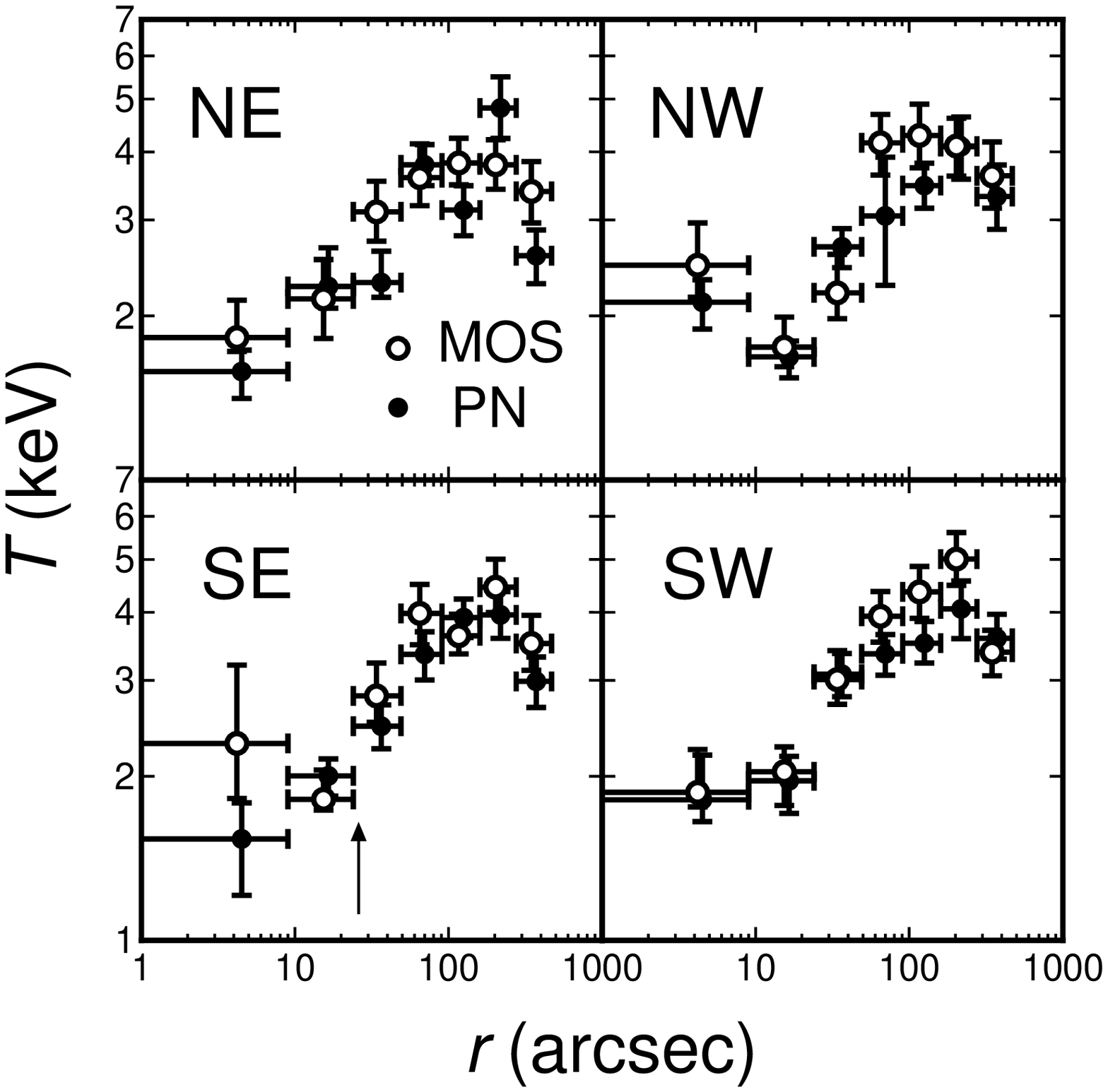} \caption{The same as Figure~\ref{fig:temp_proj} but
 for the deprojection analysis. \label{fig:temp_dproj}}
\end{figure}

The deprojection introduces anticorrelated errors. Thus, to avoid
non-physical oscillations in the temperature and abundance profiles we
employed wider radial bins than those used for the projected profiles.
In the deprojected case, the temperature drops to a lower value at the
center of the cluster than in the projected case
(Fig.~\ref{fig:temp_dproj}), because the projected hotter gas raises the
apparent average temperature. The temperature jump seen in
Figure~\ref{fig:temp_proj} is also seen in Figure~\ref{fig:temp_dproj}.

In the central region, the deprojected abundances are larger than the
projected ones, because the projected gas with small abundance reduces
the apparent average abundance. A further reason for this could be that
the abundance is often underestimated if an isothermal model is fitted
to a multiphase spectrum including a few thermal components
\citep[e.g.][]{buo00}. The central values are $Z\gtrsim 1\: Z_\sun$,
which are larger than those for the projected values ($Z\sim 0.8\:
Z_\sun$).  We did not find jumps in the abundance profiles even at the
temperature jump, though this may be due to larger error bars.

\subsection{Spectra of the Tongue}
\label{sec:spectra_tongue}

We would like to determine the spectrum of the X-ray tongue, correcting
for the emission from ambient cluster gas in front of or behind the
tongue but seen in projection against the tongue.  Of course, the
correction for projected emission depends on the geometry of the tongue
and the ambient gas.  We assume that the ambient gas is spherically
symmetric about the cluster center, except for the presence of the
tongue.  We further assume that the tongue does not extend in any
direction to larger radii than its outermost projected radius
(49\arcsec).  The projected ambient emission from gas at larger radii
than this is taken directly from the spectral observations at larger
radii. We did not fit the MOS and PN spectra simultaneously because in
this deprojection analysis, the emission from the tongue region consists
of that from the tongue itself and those from the projected gas at
larger radii. Since the calibration of MOS and PN depend on the
detector position, the variations could collectively affect the spectrum
of the tongue region in a complicated way; the spectra of outer shells
that are determined based on the calibration valid for outer regions of
the detector affect the spectrum of an inner shell or the tongue that
should be determined based on the calibration valid for the inner region
of the detector only.

The key uncertainty in the deprojection is the extent of the tongue
along the line of sight. In Figure~\ref{fig:tongue}, we present three
simple models of this extent. Firstly, we assumed that the extent of the
tongue along the line of sight was the same as its transverse width
(Fig.~\ref{fig:tongue}a); we refer to this as the ``Column'' model.
Secondly, the line-of-sight depth of the tongue might be comparable to
its radial extent from the cluster center, which would correspond to a
cap-like structure seen end-on.  As a specific model of this, we assumed
that the tongue filled a $\pm 30^\circ$ sector centered on the tongue on
the plane of the sky (Fig.~\ref{fig:tongue}b).  We will refer to this as
the ``Wide'' model for the tongue.  Finally, the tongue might be very
narrow along the line of sight (Fig.~\ref{fig:tongue}c, the ``Thin''
model). Based on the arguments about the origin of the tongue in
\citet{fuj02}, the Column model seems the more likely geometry. The
results of the spectral fits for these models are shown in
Table~\ref{tab:tongue}. In each case, $T_1$ and $Z_1$ are the
temperature and abundance in the tongue, respectively.  The values with
subscript 2 refer to the ambient gas near the tongue.

As a more general means to determine the spectrum of the tongue without
making any detailed models for its geometry, we assume that the
projected emission from the tongue consists of projected emission from
larger radii ($ r > 49\arcsec$) determined directly from spectra at
larger radii, and a mixture of tongue and ambient emission from radii of
$24\arcsec <r< 49 \arcsec$.  Rather than specify anything about the
ambient emission in this region, we simply fit the spectrum of tongue
and ambient emission at these radii with a two-temperature model. The
results for this less restrictive ``2T'' model are shown at the end of
Table~\ref{tab:tongue}.

In all of the models, the temperature of the tongue is smaller than that
of the surrounding regions (Fig.~\ref{fig:temp_ton_dproj}), which is the
same result as that found with {\it Chandra} \citep{fuj02}. The most
likely Column model, the Thin model, and the most general 2T model all
are consistent with the temperature of the tongue region being $\sim$1.1
keV, while the ambient cluster gas at the same radius has a temperature
of about 2.5 keV.  The general consistency of the Column, Thin, and 2T
temperatures suggests that the tongue actually is narrower along the
line of sight than its radius out from the center of the cluster.
Although the values for the abundance in the tongue ($\sim$0.4 solar)
are generally smaller than those for the surrounding ICM ($\sim0.6
Z_\sun$; Fig.~\ref{fig:abun_ton_dproj}), the uncertainties in the
abundance are quite large, and the abundance in the tongue is generally
consistent with the ambient abundance within the 90\% confidence error
bars. Thus, we have no evidence for any significant difference in the
abundance of the tongue and the ambient gas.

\subsection{Spectra for the Radio Relic}
\label{sec:spectra_relic}

In order to investigate the physical conditions in the radio relic, we
extracted the X-ray spectrum from the region shown in
Figure~\ref{fig:hardness}.  This is the same as the region `R' in
Figure~10 in \citet{fuj02} that overlaps the radio relic. Based on
Figure~\ref{fig:hardness}, the X-ray spectrum appears to be somewhat
harder within the parts of the radio relic which are to the east of the
tongue.  The primary purpose of the spectral analysis is to detect or
constrain the contributions to the spectrum of a non-thermal inverse
Compton or a very hot thermal component.  Since the number of photons
from the region is not large, we simultaneously fitted the three spectra
obtained by the three detectors (MOS1, MOS2, and PN).  First, we fitted
the spectra with a single-temperature MEKAL model with a variable
absorption (1T model in Table~\ref{tab:region}).  Then, we added a
power-law component, representing the inverse Compton emission, to the
1T model (1TPL model).  In the 1TPL model, the metal abundance was fixed
to the best-fit value in the 1T model because it could not be
constrained.  An alternative explanation for any hard X-ray emission in
the radio relic region might be hot thermal gas filling the region. To
study this, we also fit a two-temperature model with a variable
absorption (2T model).  We assume that the metal abundances of the two
thermal components are the same.

The results are shown in Table~\ref{tab:region}. They are consistent
with the {\it Chandra} results \citep{fuj02}, although the uncertainties
are smaller here.  The single-temperature model (1T) does not give a
good fit.  Both the 1TPL and 2T models give better fits to the spectrum.
Their difference in the best-fit $\chi^2$ is small, and it is difficult
to decide which provides the better fit.  The $0.3$--$10$~keV flux of
the power-law component is $2.4\times 10^{-13}<F_{\rm PL}< 5.4\times
10^{-13}\:\rm ergs\; cm^{-2}\: s^{-1}$.  This is consistent with the
{\it Chandra} limit \citep[$F_{\rm PL}< 7.1\times 10^{-13}\:\rm ergs\;
cm^{-2}\: s^{-1}$;][]{fuj02}, but is an even tighter constraint. Since
the relic is near the tongue, the X-ray emission from the tongue may
affect the spectrum of the relic region because of the PSF effect
(\S\ref{sec:spectra_projected}). However, since the amount of leaked
photons from the tongue is less than 20\% and the photon energy is small
because of the low temperature of the tongue, they do not affect the hot
or nonthermal emission from the relic region. What is more serious is
the contribution of the projected hot ICM from the outer parts of the
cluster. Assuming that the cluster is spherically symmetric outside the
tongue and the relic, we found the emission from the projected hot ICM
cannot be ignored, compared with the hot or nonthermal emission from the
relic region. Thus, we regard the upper value of the flux ($F_{\rm PL} <
5.4\times 10^{-13}$ ergs cm$^{-2}$ s$^{-1}$) as a secure upper limit on
the nonthermal emission from the radio relic.

\subsection{X-ray Spectra of Possible Weak Shock Regions}
\label{sec:spectra_shock}

The cool core of A133 has a bird-like structure, with ``wings''
extending toward north and west (Figs.~\ref{fig:image}b
and~\ref{fig:shock}a). The northwest edges of the wings form an arc (the
curve AA$'$). Based on the hardness ratio map (Fig.~\ref{fig:shock}b),
the temperatures in the wings may be somewhat higher than those of the
neighboring faint regions (on the northwest side of the curve AA$'$).
Thus, the curve AA$'$ may be a weak shock.  Based on the hardness ratio
map and the X-ray image, it appears that the shock is propagating from
the southeast to the northwest. However, the increase in the hardness
ratio around the curve AA$'$ is fairly subtle. Moreover, if a spectrum
consists of more than one temperature components, the relation between
the hardness ratio and temperature is very complicated. Thus, in order
to assess the existence of a possible temperature jump, we ``measured''
the temperatures by spectral fitting.

First, we investigated whether the shock is passing through the core and
tongue. We chose three regions, To1, To2, and To3, shown in
Figure~\ref{fig:shock}. We fitted the spectra of each region obtained 
with MOS1, MOS2, and PN simultaneously because the photon counts from
the regions are not large ($N\lesssim 1000$ for MOS1 and MOS2, and
$\lesssim 3000$ for PN).  We initially assumed one thermal component
(MEKAL) with fixed Galactic absorption \citep[$1.58\times 10^{20}\rm\:
cm^{-2}$;][]{sta92}, and we refer to this model as `1T'.

The results of the fits are shown in Table~\ref{tab:tongue_sh}.  There
are no significant differences in the temperatures between the expected
upstream (To1 and To2) and downstream (To3) regions of the possible
shock.  However, the temperatures may be affected by projected hot ICM
from the outer regions of the cluster.  Thus, we added a hot thermal
component representing the projected ICM to the 1T model. For this
2T$\rm_A$ model, we assume that the metal abundances of the two thermal
components are the same because we cannot constrain the abundances
without this assumption.  We expect that the cooler component ($T_1$)
represents the gas of the tongue. The temperature $T_1$ may rise by a
factor of $1.5_{-0.3}^{+0.4}$ from region To2 to To3. In the 2T$\rm_B$
model, we forced the temperatures of the projected hotter component and
the abundances to be the same for To1, To2, and To3. We assumed that the
temperatures are 3.6~keV and the abundances are 0.7~$Z_\sun$, which are
the averages of those for the 2T$\rm_A$ model. In this model, the error
of the increase of the temperature $T_1$ from region To2 to To3 becomes
smaller and the increase is a factor of $1.3_{-0.2}^{+0.0}$.

\begin{figure}\epsscale{1.0}
\plotone{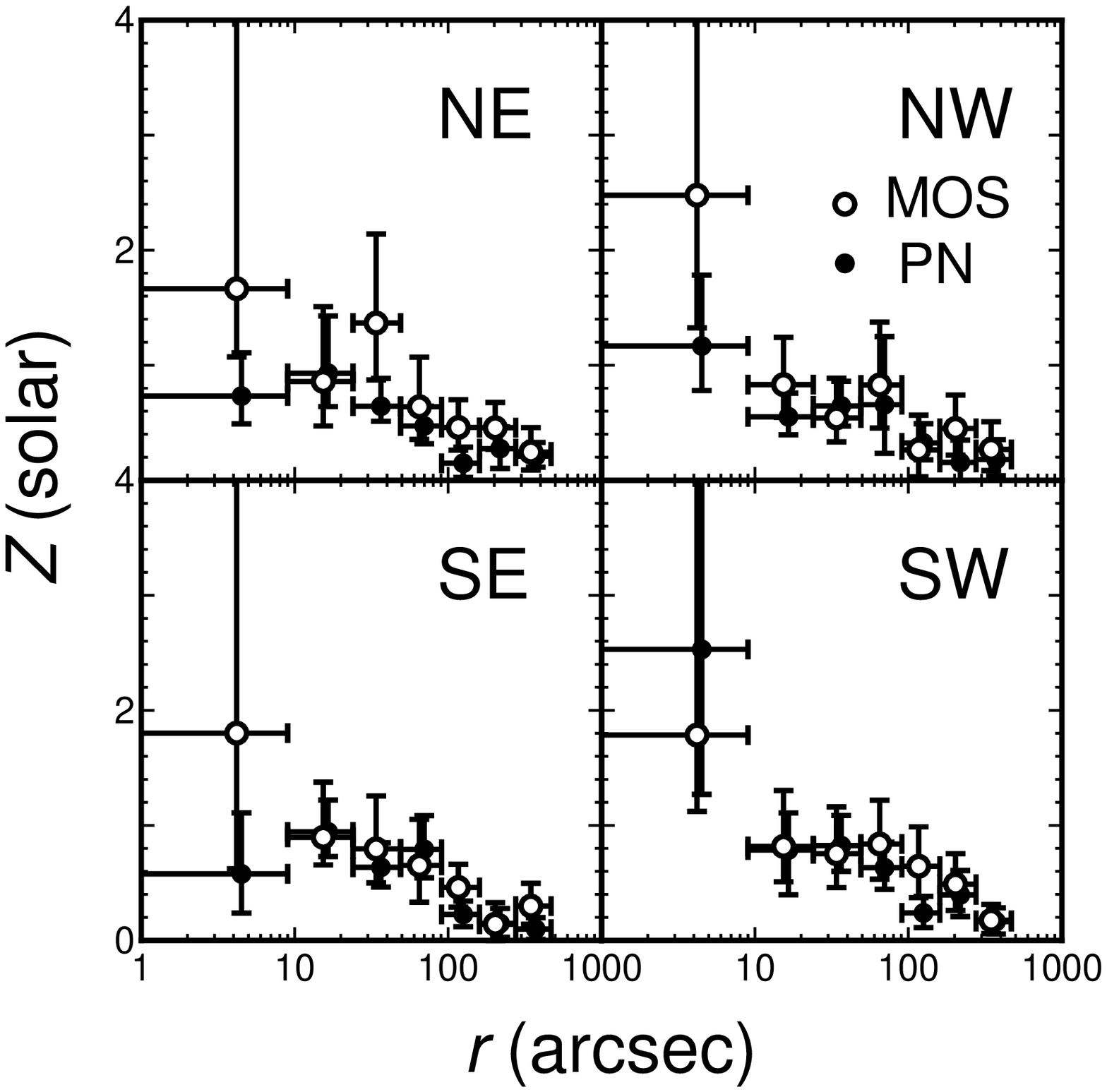} \caption{The same as Figure~\ref{fig:abun_proj} but
 for the deprojection analysis. \label{fig:abun_dproj}}
\end{figure}

\begin{figure}\epsscale{1.0}
\plotone{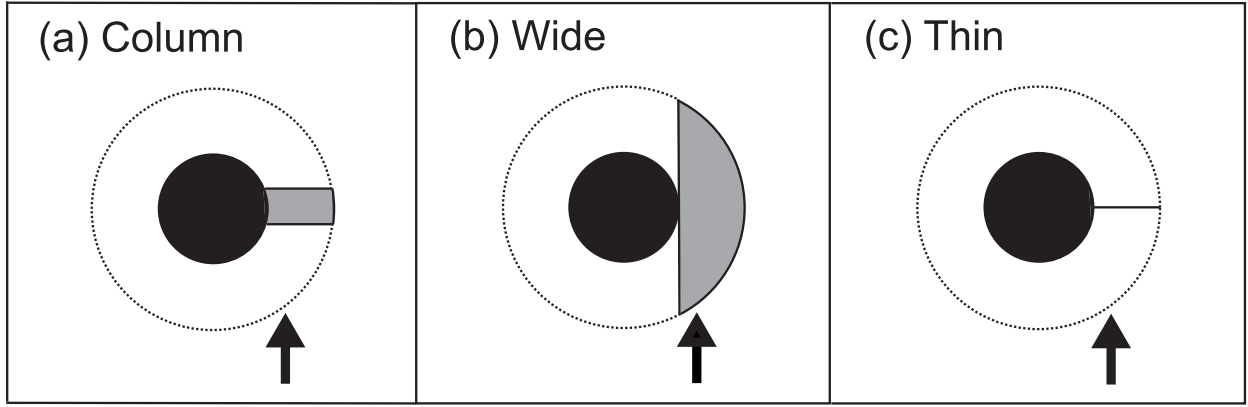}
 \caption{Possible geometries of the tongue. The arrows
 indicate the line of sight. Gray areas indicate the tongue and 
 black areas indicate the cool core as seen perpendicular to our line
 of sight.
 (a) Column model.
 (b) Wide model.
 (c) Thin model.
 \label{fig:tongue}}
\end{figure}

\begin{figure}\epsscale{1.0}
\plotone{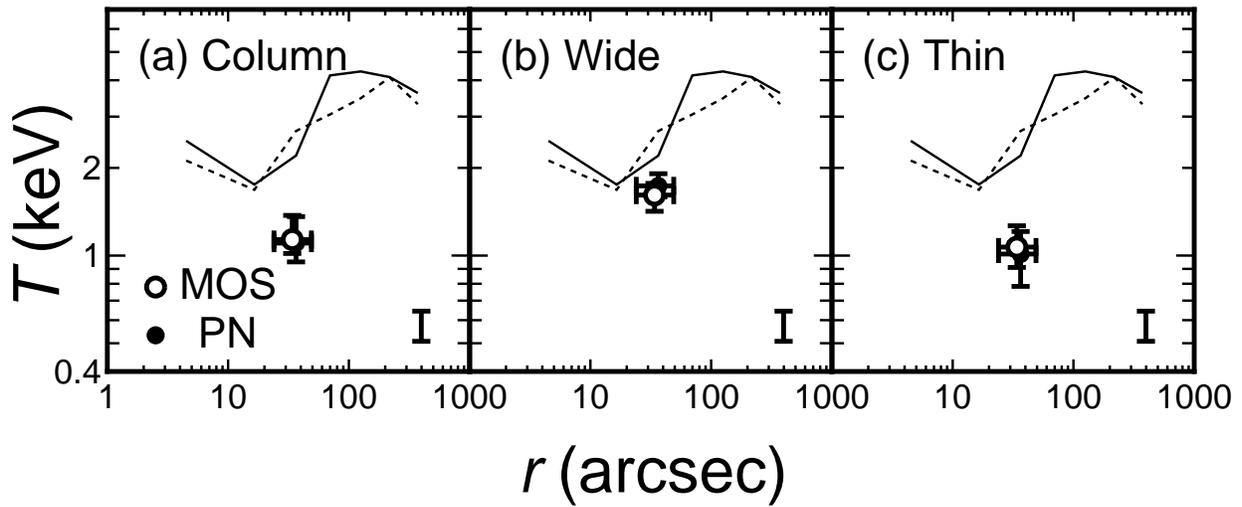}
 \caption{The same as Figure~\ref{fig:temp_ton_proj} but
 for the deprojection analysis.
 (a) Column model.
 (b) Wide model.
 (c) Thin model.
 The typical errors for the lines are shown by the bars at the lower right
corners. \label{fig:temp_ton_dproj}}
\end{figure}

\begin{figure}\epsscale{1.0}
\plotone{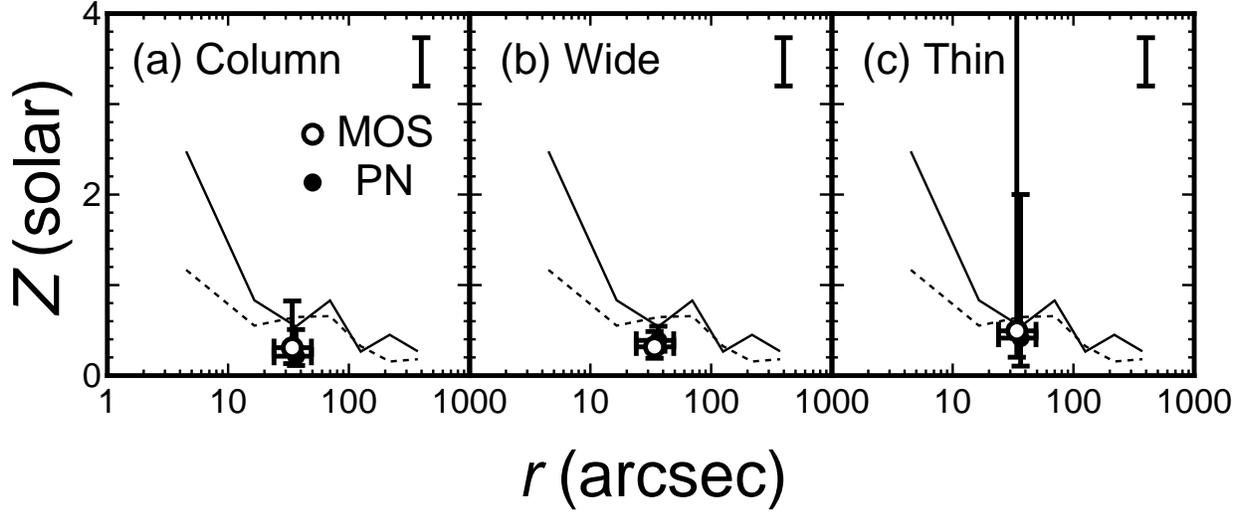}
 \caption{The same as Figure~\ref{fig:temp_ton_dproj} but
 for metal abundance.
 The typical errors for the lines are shown by the bars at the upper right
corners.  \label{fig:abun_ton_dproj}}
\end{figure}

\begin{figure}\epsscale{0.9}
\plottwo{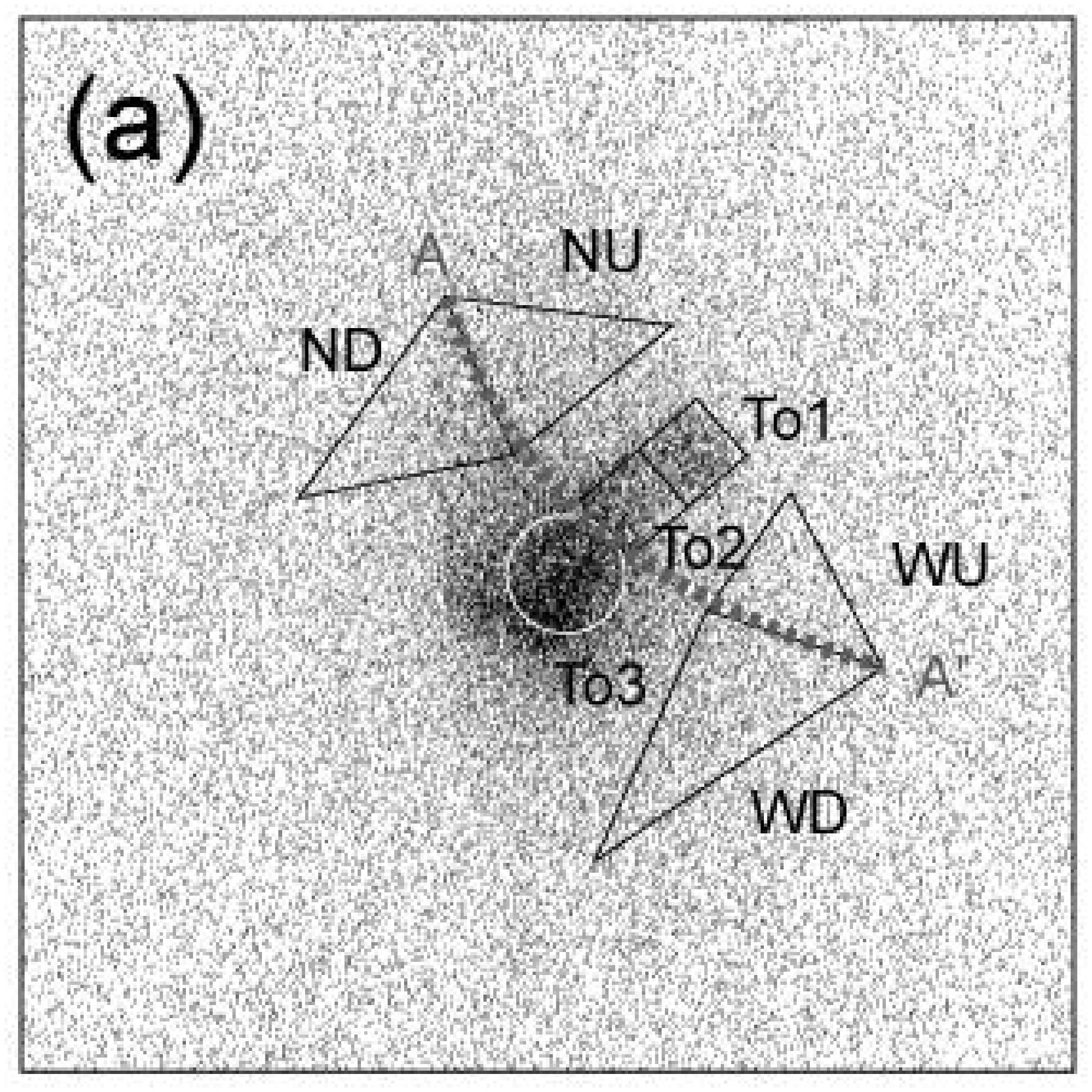}{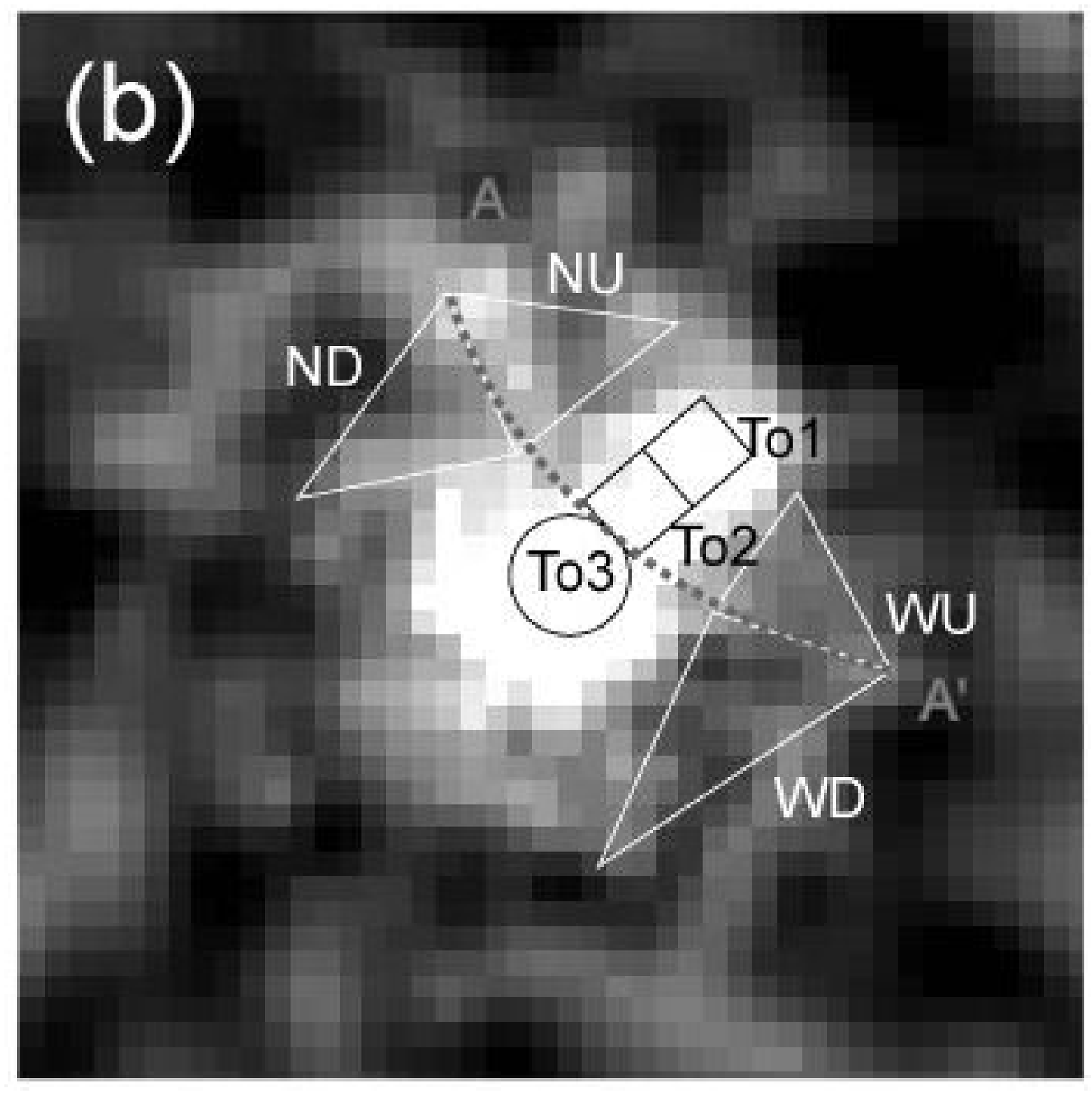} \caption{Regions for spectral analysis of
 the possible weak shock are overlaid on (a) a raw {\it Chandra} image
 and (b) the hardness ratio map (Fig.~\ref{fig:hardness}), where hard
 emission appears black. The image sizes are $3'\times 3'$. The curve
 AA$'$ is a possible weak shock.  The regions ND, NU, WD, WU are the
 north downstream, north upstream, west downstream and west upstream in
 the simple shock model presented.  On the cluster core and tongue, To1
 and To2 are upstream, while To3 is downstream.  \label{fig:shock}}
\end{figure}

We also investigated the spectra of four wing regions, WU, WD, NU, and
ND (west upstream, west downstream, etc.) shown in
Figure~\ref{fig:shock}. The curve AA' corresponds to the small gap of
the surface brightness seen on the {\it Chandra} image
(Fig.\ref{fig:image}b). The four wing regions were chosen so that the
temperature difference between the upstream and downstream regions was
the largest by looking at the hardness ratio map
(Fig.~\ref{fig:shock}b). Moreover, we selected these regions because
they are sufficiently far from the tongue that they are not affected by
the tongue's emission. Again, we fitted the spectra of each region
obtained with MOS1, MOS2, and PN simultaneously.  The results are
presented in Table~\ref{tab:shock}.  For the single-temperature 1T
model, there are no significant differences in the temperatures between
the expected upstream (NU, WU) and downstream (ND, WD) regions of the
possible shock.  However, since the projection of the hot ICM in the
outer region of the cluster may affect the results, we added another hot
thermal component representing the projected ICM to the 1T models.
Since the photon counts in these regions are particularly small, the
temperature and metal abundance of the projected ICM were fixed at
$T=4.0$~keV and $Z=0.6\: Z_\sun$, which are the average values of the
ICM in the outer region of the cluster ($50\arcsec\lesssim r \lesssim
100\arcsec$; Figs.~\ref{fig:temp_proj}, \ref{fig:abun_proj},
\ref{fig:temp_dproj} and~\ref{fig:abun_dproj}). Without fixing the
projected component, we cannot constrain the cooler component. In order
to reduce the number of free parameters further, we also fixed the
abundance of the cooler component at $Z=0.6\: Z_\sun$.  We fitted the
spectra of the expected upstream (WU and NU) and downstream (WD and ND)
regions with the two thermal components.  The free parameters in this 2T
model are the temperature and normalization of the cooler component and
the normalization of the hotter component.  However, while the
temperature and normalization of the cooler component are independent in
different regions, we adjusted the normalization of the hotter $4.0$~keV
component in the downstream regions to that in the upstream regions in
such a way that the normalization of the $4.0$~keV component in the WD
(ND) region is assumed to be that in the WU (NU) region multiplied by
the ratio of the area of WD to that of WU (the area of ND to that of
NU). This means that the spectral contribution from the projected hotter
component per unit area is the same for the WD (ND) and WU (NU)
regions. The results are shown in Table~\ref{tab:shock}. The spectra of
the upstream and downstream regions are fitted simultaneously contrary
to the 1T models. Therefore, for the 2T models, the degree of freedom
and $\chi^2$ of the WD and WU (ND and NU) are the same and much larger
than those for the 1T models (Table~\ref{tab:shock}). For the cooler
component in the west wing, the temperature rises from regions WU to WD
by a factor of $2.3\pm 0.6$. For the north wing, we did not find
evidence for a temperature change.

\section{Discussion}

\subsection{Cold Front}
\label{sec:coldfront}

The SE temperature jump found by us corresponds to the jump in surface
brightness as seen in {\it Chandra} data \citep{fuj02}. Deprojection
analysis showed that the density increases there by a factor of 1.3 from
the outer to the inner region \citep[Fig.~7 in][]{fuj02}. On the other
hand, the temperature decreases by a factor of $1.4$ from the outer to
the inner region. If we average the MOS and PN data about the
temperature jump, the ratio of the pressures on the sides of the jump is
$0.95\pm 0.11$, which is consistent with a continuous pressure across
the jump. Thus, this jump may be classified as a ``cold front'', and is
similar to the one found in A1795 \citep*{mar01}.  From equation~(2) in
\citet{vik01}, we found an upper limit of $U\le 230\:\rm km\: s^{-1}$
for the relative velocity. In this estimation, we assumed the density
and the temperature just inside the jump to be the same as those at the
contact discontinuity.  We estimated the gas pressure far upstream from
the cold front by extrapolating the density and temperature inward from
those determined at radii $r\sim 40-60''$ in the direction of the cold
front.

\subsection{Weak Shock}

A detailed spectral analysis showed that the bird-like feature of the
cool core seems to be a weak shock (\S\ref{sec:spectra_shock}). In fact,
the direction of the bend of the curve AA$'$ is consistent with this
idea. When a shock passes the cool core from the southeast to the
northwest, the velocity of the central part of the shock that passes
through the cool core is smaller than the velocity of the part of the
shock that passes around the core, because of the larger pressure and
lower temperature of the core \citep[See Fig.~11 in][]{chu03}.  It is
this differential velocity that bends the shock.

In the core and tongue, the temperature rises from the region To2 to~To3
by a factor of $1.3_{-0.2}^{+0.0}$ in the model 2$\rm T_B$
(Table~\ref{tab:tongue_sh}). The Mach number of the shock is
$1.3_{-0.2}^{+0.0}$ as derived from the Rankine-Hugoniot relation:
\begin{equation}
 \frac{T_d}{T_u}=\frac{[2\gamma M_u^2-(\gamma-1)][(\gamma-1)M_u^2+2]}
  {(\gamma+1)^2 M_u^2} \:,
\end{equation}
where $T_u$ is the upstream temperature, $T_d$ is the downstream
temperature, $\gamma (=5/3)$ is the adiabatic index, and $M_u$ is the
upstream Mach number. Thus, the shock is marginally detected. On the
other hand, the temperature ratio at the western shock is $2.3\pm 0.6$
(Table~\ref{tab:shock}), and the Mach number of the shock is
$1.6$--$2.6$ from the Rankine-Hugoniot relation. However, the evidence
of the western shock is based on keeping the temperatures of the
projected hot component fixed. There is no such evidence when all
temperatures are left unrestricted. (\S\ref{sec:spectra_shock}).

The shape of the shock front shows that the shock is not directly
related to the central AGN of the cluster (Fig.~\ref{fig:image}b).
Thus, it may be different from the weak shocks found in the core of the
Perseus cluster \citep{fab03}. It is more possible that the shock
originated outside the core region and propagated into the core. If so,
this is the first time that a sign of this kind of shock is found in a
cluster core. The shock might have formed in a former cluster merger.
Alternatively, it might have formed through steepening of the fronts of
sound waves with large amplitudes produced by ICM motion in the
cluster. This is because the sound speed of the compressed gas is higher
than that in the rarefied part, so the waves will steepen over a few
wavelengths \citep[e.g. Fig.\ 55.1 in][]{mih84}. Recently,
\citet*{fuj04a} showed analytically and numerically that sound waves
formed outside a cluster core steepen and become weak shocks in the
core. These shocks heat the cool core and increase the cooling time of
the gas in the core, which may solve the cooling flow problem.  Some of
the complicated X-ray structures in the core found in the {\it Chandra}
observations (Fig.~\ref{fig:image}b) may be produced by the passage of
this shock \citep*{fuj04b}. We note that the regions around the
curve~AA$'$ we took for the spectral analysis are fairly
large. Therefore, the temperature jump we found may not be as sharp as
expected for a shock. In this case, the gradient of the temperature and
surface brightness could be explained by a sound wave. Since the
amplitude represented by the temperature gradient is relatively large,
the wave would become a weak shock as it propagates only a few
wavelengths \citep{mih84}.

\subsection{The Origin of the X-ray Tongue}

\citet{fuj02} discussed several models for the origin of the tongue: (1)
a cooling wake; (2) a cluster merger; (3) Kelvin-Helmholtz (KH)
instabilities; or (4) a buoyant radio bubble.  In the cooling-wake
scenario the tongue is the gas cooling from the hot ICM, attracted into
a wake along the path of the moving cD galaxy \citep{dav94,fab01}. The
cluster-merger scenario predicts that the tongue is formed through
convection induced by a cluster merger \citep{ric01}. In the
KH-instability scenario, the cool core is moving relative to the
surrounding hot ICM. As a result, KH instabilities develop around the
core and produce the tongue.  In the buoyant bubble scenario, the radio
relic is an old radio lobe, which is moving outward in the cluster
gravitational potential. This buoyant radio bubble lifts up cooler gas
from the cluster center \citep[e.g.][]{chu01,qui01,bru01}.  In the
following, we compare each of these scenarios using our combined {\it
XMM-Newton} and {\it Chandra} data.

\subsubsection{A Cooling Wake}
\label{sec:wake}

\citet{fuj02} concluded that the cooling-wake scenario is not favored by
the steep temperature gradient on both sides of the tongue; the
temperature should be continuous if the cold gas of the tongue is a
result of the cooling flow from the surrounding ICM.  The {\it
XMM-Newton} results are consistent with this, albeit with poorer spatial
resolution. However, the {\it XMM-Newton} observations, with their
information on the metal abundance, may provide another clue. If the
metal abundance around the tongue were also discontinuous, it would
strengthen the previous conclusion and make the cooling wake model even
more unlikely. Table~\ref{tab:tongue} and
Figure~\ref{fig:abun_ton_dproj} show that, although the best-fit
abundance in the tongue is generally somewhat lower than that in the
ambient gas, the two are consistent within the uncertainties.  Thus,
based only on the metal abundance, we cannot reject the cooling-wake
scenario.

\subsubsection{Cluster Merger}
\label{sec:merger}

\citet{fuj02} also concluded that the cluster merger scenario is not
suitable to explain the origin of the tongue since no clear evidence of
a shock was seen, neither in the image, nor in the X-ray hardness map,
nor in the {\it Chandra} spectra.  However, by comparing the {\it
XMM-Newton} and {\it Chandra} observations, we have found a possible
weak shock in the cool core (\S\ref{sec:spectra_shock}). Moreover, if
the ratio of the masses of the merging clusters is large (minor merger),
we may not necessarily expect to find clear structures, such as a strong
shock, associated with the merger.  \citet{gom02} performed numerical
simulations of the head-on merger of a cooling flow cluster with an
infalling subcluster of galaxies. They adopted relatively high mass
ratios of 16:1 and 4:1. Since they included the effect of gas cooling,
the central density of the cluster is high. Because of this, in some
cases the gas of the subcluster cannot penetrate the cooling core of the
larger cluster, while the dark matter can. As a result, the cool core of
the cluster is not destroyed by the merger.  Well after the merger
($\gtrsim 1$~Gyr), the cluster becomes spherical \citep{gom02}. However,
the oscillation of the ICM may elongate the core and produce a
tongue-like structure \citep[Fig.~6 in][]{gom02}.  Moreover, the
sloshing of the ICM around the core could produce the temperature jump
at the southeast edge of the core, and a temperature jump of this type
may exist without a pressure discontinuity \citep{mar01,chu03}. Thus, if
the mass ratio of the merging clusters is large, a cluster merger
remains a candidate for the origin of the tongue and the temperature
jump.

\subsubsection{Kelvin-Helmholtz Instability}

The KH instability scenario seemed plausible in \citet{fuj02}.  However,
these authors could not estimate the velocity of the cool core from {\it
Chandra} data, and thus they assumed it was sufficient to produce the
instability.  In \S~\ref{sec:coldfront} of the present paper, we derived
an upper limit on the velocity of the cool core based on the {\it
XMM-Newton} data.  The minimum velocity for the development of KH
instabilities is,
\begin{equation}
\label{eq:KHU}
 U_{\rm KH} \equiv
\sqrt{\frac{k_B T_{\rm in}}{\pi \mu m_H} \frac{D^2-1}{D}}
=230\rm\: km\: s^{-1} \, ,
\end{equation}
where $k_B$ is the Boltzmann constant, $T_{\rm in}$ is the
temperature on the side of the temperature jump which is closer to the
cluster center, $m_H$ is the mass of the hydrogen atom, and $D\; (\geq
1)$ is the density contrast at the edge of the cool core \citep{fuj02}.
On the other hand, we found the velocity of the cool core is $U<230\rm\:
km\: s^{-1}$ (\S\ref{sec:coldfront}), which is smaller than $U_{\rm
KH}$. This means that KH instabilities are unlikely to develop around
the cool core.

\subsubsection{Buoyant Radio Bubble}
\label{sec:bubble}

{\it Chandra} data had shown that there is an X-ray deficit in the
region around the radio relic northwest of the cluster center
\citep{fuj02}.  Thus, there seems to be a bubble in the ICM. If the
tongue is the gas uplifted by the buoyant bubble, the properties of its
gas should be the same as those at the cluster center. The metal
abundance at the cluster center is $Z\gtrsim 1.0\: Z_\sun$
(Fig.~\ref{fig:abun_dproj}). As discussed in \S\ref{sec:wake}, although
the best-fit abundance in the tongue is lower, the uncertainties are
large enough to be consistent with those at the cluster center, at least
if the volume of the tongue is small
(Fig.~~\ref{fig:abun_ton_dproj}c). Thus, this scenario is still
plausible.

\citet{sle01} estimated that the age of the relic is $t_{\rm
radio}=4.9\times 10^7$~yr. If the cD galaxy were located at the position
of the relic when the relic was formed, the velocity of the galaxy must
be $U \gtrsim 700\:\rm km\: s^{-1}$ \citep{fuj02}. This is inconsistent
with the velocity of $U< 230\rm\: km\: s^{-1}$ inferred from the present
observations. However, buoyancy provides a viable transport mechanism
for the relic, since the required velocity of the bubble is consistent
with the results of numerical simulations \citep[e.g.,][]{chu01}.

\subsection{The Radio Relic}

We found an upper limit on the nonthermal X-ray emission from the radio
relic of $F_{\rm PL} < 5.4\times 10^{-13}$ ergs cm$^{-2}$ s$^{-1}$. From
this limit, we can derive a lower limit on the magnetic field in the
relic by comparing this limit to the inverse Compton emission expected
from the radio-emitting electrons.  We adopt an X-ray photon index of
1.8 and a radio flux density of $35.5\pm 4.3$~Jy at 80~MHz
\citep{sle01}.  Using the standard expressions for the inverse Compton
and synchrotron emission assuming a power-law electron energy
distribution \citep[e.g.,][]{sar86}, we can constrain the required
magnetic field in the relic as $B\geq 1.5$ $\mu$G, which is consistent
with the {\it Chandra} result \citep[$B\geq 1.2$ $\mu$G;][]{fuj02}. This
value is at least consistent with the magnetic field of $14.4\: \mu$G
derived from minimum-energy arguments \citep{sle01}.

\section{Conclusions}

We have presented the results of an {\it XMM-Newton} observation of the
cluster of galaxies A133. We found a probable cold front to the
southeast of the cool core.  The pressure around the front shows that
the relative velocity of the core with respect to the rest of the
cluster is very small ($U< 230\rm\: km\: s^{-1}$). This cold front was
first identified as a surface brightness discontinuity with {\it
Chandra}, but only with {\it XMM-Newton} spectra were we able to
determine the nature of this feature.

The {\it Chandra} image of A133 showed a complex, bird-like morphology
in the cluster core. Based on the {\it XMM-Newton} spectra and hardness
ratio maps, we argue that the wings of this structure are a weak shock
front. However, this evidence is marginal as it is based keeping the
temperatures of the projected hot component fixed, and the evidence
disappears when all temperatures are left free. The shock appears to be
propagating from the southeast to the northwest.  Its curved shape
suggests that it has already passed through the cluster core, and that
the central portion of the shock was retarded by the higher density at
the cluster center.  This morphology is not consistent with the shock
arising from the active nucleus of the cluster's central cD galaxy,
unlike the weak shocks which have been seen in the {\it Chandra} image
of the Perseus cluster \citep{fab03}.  To our knowledge, the morphology
and origin of this shock are new features which have not been observed
in other clusters so far.  This shock may be heating the cluster core.

Our previous {\it Chandra} observations showed a ``tongue'' of
relatively cool gas extending from the center of the cD to the center of
the radio relic.  We have used {\it XMM-Newton} spectra to test various
models for the origin of this tongue.  Both the {\it Chandra} and {\it
XMM-Newton} observations indicate that the temperature in the tongue
drops discontinuously at its sides, which is inconsistent with it being
a cooling wake.  A discontinuity of the metal abundance would be a
further argument against this origin, but the {\it XMM-Newton} spectra
do not allow any firm conclusion about such a discontinuity.  The small
velocity of the core, as determined from the pressure profile near the
cold front, is inconsistent with the possibility that the tongue is
formed through Kelvin-Helmholtz instabilities around the core.  The
tongue may well be gas which has been uplifted by a buoyant radio bubble
including the radio relic in the northwest of the core.  This model
predicts that the metal abundances in the tongue be high and comparable
to those at the very center of the cluster, where the gas would have
originated. Despite the fact that the best-fit abundances in the tongue
derived from {\it XMM-Newton} spectra are lower than predicted, the
uncertainties do allow for high abundances as well.  Thus, a buoyant
bubble is still a promising candidate for the origin of the tongue.  Our
{\it XMM-Newton} observation has revived interest in the possibility
that the tongue results from a cluster merger.  Previously, the main
argument against this model was the lack of evidence for any
merger-related shocks.  Our discovery of a likely weak merger shock
supports this model, which probably requires a very unequal merger with
a large mass ratio between the main cluster and the merging subcluster.

There is no clear evidence of nonthermal inverse Compton emission from
the radio relic in the X-ray band.  The upper limit on the inverse
Compton contribution based on our {\it XMM-Newton} spectra implies a
lower limit on the magnetic field in the radio relic of $B\geq
1.5$~$\mu$G.

\acknowledgments

We would like to thank E.~L. Blanton, T. Tamura, and D. Balsara for
helpful discussions. Support for this work was provided by the National
Aeronautics and Space Administration through {\it XMM-Newton} Award
Number NAG5-13088 and NAG5-13737, and Chandra Award Number GO2-3159X,
issued by the $Chandra$ X-ray Observatory Center, which is operated by
the Smithsonian Astrophysical Observatory for and on behalf of NASA
under contract NAS8-39073.  Y.~F.\ was supported in part by a
Grant-in-Aid from the Ministry of Education, Culture, Sports, Science,
and Technology of Japan (14740175). T.~H.~R. acknowledges support by the
F.H. Levinson Fund of the Peninsula Community Foundation through a
Post-Doctoral Fellowship. H.~A. acknowledges support from CONACyT grant
40094-F.  At the University of Minnesota, this research is supported, in
part, by National Science Foundation Grant AST-0307600.  This work is
based on observations obtained with {\it XMM-Newton} an ESA science
mission with instruments and contributions directly funded by ESA Member
States and the USA (NASA).

\clearpage

\begin{deluxetable}{cccccc}
\tabletypesize{\footnotesize}
\tablecaption{Deprojected Spectra of the Tongue\label{tab:tongue}}
\tablewidth{0pt}
\tablehead{
\colhead{} & \colhead{$T_1$} & \colhead{$T_2$} 
& \colhead{$Z_1$}   & \colhead{$Z_2$}   & 
  \colhead{$\chi^2$/dof} \\
\colhead{Model}  & \colhead{(keV)} & \colhead{(keV)} & 
\colhead{($Z_{\sun}$)}  & \colhead{($Z_{\sun}$)} &   
}
\startdata
Column (MOS)  & $1.1^{+0.3}_{-0.1}$  & 2.2\tablenotemark{a}
                           &$0.31^{+0.18}_{-0.51}$ & 0.54\tablenotemark{a}
                                          & 0.886 (22.2/25) \\
Column (PN) & $1.1^{+0.3}_{-0.1}$  & 2.6\tablenotemark{a}
                           &$0.21^{+0.30}_{-0.10}$ & 0.65\tablenotemark{a}
                                          & 0.950 (32.3/34) \\
Wide (MOS) &$1.6^{+0.2}_{-0.2}$ & \nodata 
                     &$0.32^{+0.17}_{-0.13}$ & \nodata & 1.11 (27.6/25) \\
Wide (PN) &$1.7^{+0.2}_{-0.1}$  & \nodata 
                    &$0.39^{+0.15}_{-0.12}$ & \nodata & 0.950 (32.3/34) \\
Thin (MOS) &$1.1^{+0.2}_{-0.2}$  & 2.2\tablenotemark{a}
	               &$0.49^{+\infty}_{-0.29}$ & 0.54\tablenotemark{a} 
	                   & 0.820 (20.5/25) \\
Thin (PN) &$1.0^{+0.2}_{-0.2}$  & 2.6\tablenotemark{a}
                           &$0.41^{+1.59}_{-0.31}$ 
                           & 0.65\tablenotemark{a} & 1.10 (37.5/34) \\
2T (MOS)  & 
$1.1^{+0.3}_{-0.3}$  & $2.6^{+1.9}_{-0.6}$
                       &$0.51^{+\infty}_{-0.32}$ & 0.54\tablenotemark{a}
                                         & 0.865 (29.9/23) \\
2T (PN)   &
$0.8^{+0.6}_{-0.3}$  & $2.0^{+0.8}_{-0.2}$
                          &$0.15^{+0.60}_{-0.12}$ & 0.65\tablenotemark{a}
                                         & 0.859 (27.5/32) \\
\enddata
\tablecomments{The cooler component ($T_1$, $Z_1$) represents the 
gas of the tongue,
and the hotter component ($T_2$, $Z_2$) represents the surrounding gas.}

\tablenotetext{a}{Fixed at the value outside the tongue.}
\end{deluxetable}


\begin{deluxetable}{ccccccc}
\tabletypesize{\footnotesize}
\tablecaption{Spectra for the Radio Relic Region
\label{tab:region}}
\tablewidth{0pt}
\tablehead{
\colhead{} & \colhead{$T_1$} & \colhead{$T_2$}
& \colhead{$Z$} & \colhead{$\Gamma$}   
& \colhead{$N_{\rm H}$} & \colhead{$\chi^2$/dof}  \\
\colhead{Model}  & \colhead{(keV)} & \colhead{(keV)} & 
\colhead{($Z_{\sun}$)}     & \colhead{}  &
\colhead{($\times 10^{20}\rm\; cm^{-2}$)} & \colhead{}
}
\startdata
 1T   &$2.8_{-0.1}^{+0.1}$&\nodata 
                  &$0.72_{-0.12}^{+0.15}$&\nodata 
                       &$0.93_{-0.64}^{+0.67}$ & 1.18 (189.6/160) \\
 1TPL &$2.1_{-0.2}^{+0.1}$
                  &\nodata & 0.72\tablenotemark{a} 
                                           &$1.8_{-0.2}^{+0.1}$ 
                                           &$2.51_{-1.15}^{+1.38}$ 
                                                 & 1.05 (167.0/159)\\
  2T   &$1.6_{-0.5}^{+0.6}$&$3.6_{-0.6}^{+\infty}$ 
                           &$0.65_{-0.27}^{+0.23}$ &\nodata 
                                           &$0.74_{-0.8}^{+0.6}$ 
                                                 & 1.06 (167.7/158) \\
\enddata
\tablenotetext{a}{Fixed at the value from the 1T model.}
\tablenotetext{b}{Fixed.}
\end{deluxetable}

\begin{deluxetable}{cccccc}
\tabletypesize{\footnotesize}
\tablecaption{Spectra of the Core and Tongue\label{tab:tongue_sh}}
\tablewidth{0pt}
\tablehead{
\colhead{} & \colhead{} & \colhead{$T_1$} & \colhead{$T_2$} 
& \colhead{$Z$}   & 
  \colhead{$\chi^2$/dof} \\
\colhead{Region}  & \colhead{Model}  & \colhead{(keV)} 
& \colhead{(keV)} & \colhead{($Z_{\sun}$)}  & 
}
\startdata
To1 & 1T & $1.6^{+0.1}_{-0.1}$ & \nodata 
                          &$0.38^{+0.12}_{-0.09}$ &  1.89 (68.1/36) \\
To1 & 2T$\rm_A$ & $1.1^{+0.3}_{-0.2}$ & $2.4^{+\infty}_{-0.4}$
                          &$0.57^{+0.40}_{-0.33}$ &  1.38 (46.8/34) \\
To1 & 2T$\rm_B$ & $1.3^{+0.1}_{-0.1}$ & 3.6\tablenotemark{a} 
                          &0.7\tablenotemark{a}   &  1.54 (55.3/36) \\
To2 & 1T & $2.0^{+0.1}_{-0.2}$ & \nodata
                          &$0.64^{+0.17}_{-0.14}$ &  1.51 (80.0/53) \\
To2 & 2T$\rm_A$ & $1.1^{+0.3}_{-0.1}$ & $2.8^{+0.8}_{-0.4}$
                          &$0.81^{+0.31}_{-0.18}$ &  1.00 (51.2/51) \\
To2 & 2T$\rm_B$ & $1.3^{+0.1}_{-0.1}$ & 3.6\tablenotemark{a} 
                          &0.7\tablenotemark{a}   &  1.00 (53.0/53) \\
To3 & 1T &$2.0^{+0.1}_{-0.1}$  & \nodata
                          &$0.75^{+0.12}_{-0.08}$ & 1.41 (205.9/146) \\
To3 & 2T$\rm_A$ & $1.7^{+0.1}_{-0.3}$  & $5.6^{+\infty}_{-2.4}$
                          &$0.71^{+0.16}_{-0.11}$ 
                          & 1.30 (187.0/144) \\
To3 & 2T$\rm_B$ & $1.7^{+0.1}_{-0.1}$ & 3.6\tablenotemark{a} 
                          &0.7\tablenotemark{a}   &  1.29 (187.6/146) \\
\enddata
\tablenotetext{a}{Fixed.}
\end{deluxetable}


\begin{deluxetable}{ccccc}
\tabletypesize{\footnotesize}
\tablecaption{Spectra around the Weak Shock \label{tab:shock}}
\tablewidth{0pt}
\tablehead{
\colhead{} & \colhead{} & \colhead{$T$} & \colhead{$Z$}   & 
  \colhead{$\chi^2$/dof} \\
\colhead{Region} & \colhead{Model}  & \colhead{(keV)} & 
\colhead{($Z_{\sun}$)}     &   
}
\startdata
WU & 1T &$3.2_{-0.4}^{+0.4}$ 
                          &$0.35_{-0.19}^{+0.31}$ & 0.955 (33.4/35) \\
   & 2T &$0.9_{-0.2}^{+0.2}$ 
                          &0.6\tablenotemark{a}  & 0.990 (122.7/124) \\
WD & 1T &$3.2_{-0.2}^{+0.2}$ 
                          &$0.61_{-0.15}^{+0.17}$ & 1.08 (95.8/88) \\
   & 2T &$2.0_{-0.2}^{+0.4}$ 
                          &0.6\tablenotemark{a}  & 0.990 (122.7/124) \\
NU & 1T &$2.7_{-0.2}^{+0.2}$ 
                          &$0.50_{-0.14}^{+0.17}$ & 1.50 (102.0/68) \\
   & 2T &$2.2_{-0.5}^{+0.2}$ 
                          &0.6\tablenotemark{a}  & 1.12 (157.0/140) \\
ND & 1T &$2.7_{-0.2}^{+0.2}$ 
                          &$0.62_{-0.15}^{+0.19}$ & 1.06 (75.1/71) \\
   & 2T &$2.1_{-0.4}^{+0.5}$ 
                          &0.6\tablenotemark{a}  & 1.12 (157.0/140) \\
\enddata
\tablecomments{The shock region temperatures and abundances in the 2T
model are those of the cooler component. The 2T fits have much more
degrees freedom than the 1T fits because the 2T fits are simultaneous
fits of the upstream and downstream regions whereas 1T fits are
individual fits (see text).}  
\tablenotetext{a}{Fixed.}
\end{deluxetable}

\clearpage

\clearpage

\clearpage

\clearpage

\clearpage

\end{document}